\documentclass[aps,prb,twocolumn,showpacs,superscriptaddress]{revtex4}

\usepackage{graphicx}
\usepackage{amsmath}

\begin{document}


\title{Determination of magnetic order of the rare-earth ions in multiferroic $\mathbf{TbMn_2O_5}$}



\author{R. D. Johnson}
\author{S. R. Bland}
\affiliation{Department of Physics, Durham University, Rochester Building, South Road, Durham, DH1 3LE, United Kingdom}
\author{C. Mazzoli}
\affiliation{ESRF, 6 rue Jules Horowitz, BP220, 38043 Grenoble Cedex, France}
\author{T. A. W. Beale}
\affiliation{Department of Physics, Durham University, Rochester Building, South Road, Durham, DH1 3LE, United Kingdom}
\author{C-H. Du}
\affiliation{Department of Physics, Tamkang University, Tamsui 251, Taiwan}
\author{C. Detlefs}
\affiliation{European Synchrotron Radiation Facility, F-38043 Grenoble, France}
\author{S. B. Wilkins}
\affiliation{Department of Condensed Matter Physics and Materials
Science, Brookhaven National Laboratory, Upton, New York, 11973-5000, USA}
\author{P. D. Hatton}
\email{p.d.hatton@durham.ac.uk}
\affiliation{Department of Physics, Durham University, Rochester Building, South Road, Durham, DH1 3LE, United Kingdom}


\date{\today}

\begin{abstract}
	We have employed resonant x-ray magnetic scattering to specifically probe the magnetic order of the rare-earth ions in multiferroic $\mathrm{TbMn_2O_5}$. Two energy resonances were observed, one originated from the E1-E1 dipolar transition and the other from the E2-E2 quadrupolar transition. These resonances directly probe the valence 5d band and the partially occupied 4f band, respectively. First, full polarization analysis, which is a measurement of the scattered polarization as a function of incident polarization, confirmed a spin polarization of the terbium valence states (probed by the E1-E1 transition) by the $\mathrm{Mn^{4+}}$ spin density in the commensurate phase. Second, full polarization analysis data were collected in the low-temperature incommensurate and commensurate phases when tuned to the E2-E2 resonance. By employing a least-squares fitting procedure, the spin orientations of the terbium ion sublattice were refined.
\end{abstract}


\maketitle

\section{INTRODUCTION}
	A small group of materials known as magnetoelectric multiferroics has been of major scientific interest for some years\cite{nhur,spaldin05,Lee:2008yq,Cheong:2007ys}. These materials display phenomena in which both magnetic and electric orders are coupled within a single phase. In particular one of the most dramatic effects has been observed in $\mathrm{TbMn_2O_5}$, a system in which a complete reversal of electric polarization is induced in applied fields of approximately 2 T.\cite{nhur} The mechanism driving the multiferroicity in such systems is yet to be fully understood; although in studies of the $\mathit{R}\mathrm{Mn_2O_5}$ series (\textit{R} = rare earth, Y or Bi),\cite{tachibana:224425} it is clear that the rare-earth ions play an important role in the magnetoelectric coupling. It has recently been shown by Koo \textit{et al.}\cite{koo:197601} that numerous magnetic orders exist in the low-temperature phases of $\mathrm{TbMn_2O_5}$, centered on $\mathrm{Mn^{3+}}$, $\mathrm{Mn^{4+}}$, and $\mathrm{Tb^{3+}}$ sites, all with the same wave vector, giving rise to a complex spin configuration. The terbium magnetic order couples with the lattice, the effect of which increases at low temperature due to an increase in the magnitude of the average terbium magnetization. It is however not possible to directly probe the magnetic moment of the rare-earth ion using neutron-diffraction or bulk transport or magnetization measurements.
	
	In the past two decades, magnetic x-ray scattering has proved to be a useful tool in measuring long-range magnetic order in single crystals. Upon tuning incident x-rays to an absorption edge, the scattered intensity from the very weak magnetic signal can be dramatically enhanced. This, coupled with the brightness of third generation x-ray sources, makes it possible to observe otherwise undetectable weak reflections from long-range magnetic, charge and orbital order.\cite{Murakami:1998fj, Wilkins:2003uq,PhysRevLett.91.167205, PhysRevLett.88.126402,PhysRevLett.88.106403} In the case of magnetic scattering at the terbium L edges, the signal is $\sim10^4$ times stronger than found off resonance. In this paper we employ full polarization analysis to investigate the polarization dependence of the scattering amplitude, with the aim of refining the direction of magnetic moments in specific terbium electronic states in $\mathrm{TbMn_2O_5}$. 
	
	Full polarization analysis is a relatively new technique, which has been shown to be capable of unravelling multipole resonances in $\mathrm{K_2CrO_4}$\cite{mazzoli:195118} and modeling competing magnetic domain contributions within $\mathrm{NpRhGa_5}$.\cite{detlefs:024425} The Appendix briefly outlines the theory behind the scattering amplitude's dependence on magnetic moment vector and polarization, providing the basis for the simulation and model refinement of the polarization dependences measured by us.

	$\mathrm{TbMn_2O_5}$ crystallizes into the space group \textit{Pbam} with lattice parameters $a=7.3251\ \mathrm{\AA}$, $b=8.5168\ \mathrm{\AA}$, and $c=5.6750\ \mathrm{\AA}$, as measured by neutron scattering\cite{A-Alonso:1997lr,chapon04,blake05}. $\mathrm{Mn^{4+}}$ and $\mathrm{Mn^{3+}}$ ions sit in octahedral and square based pyramid oxygen coordinations, respectively\cite{A-Alonso:1997lr}.
	
\begin{figure}
\includegraphics[width=6cm]{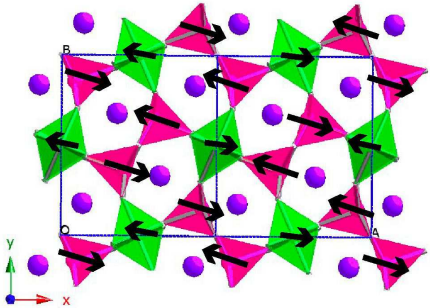}
\caption{\label{magstruc}(Color online) Illustration of the low temperature magnetic structure of $\mathrm{TbMn_2O_5}$ in the \textit{ab}-plane, adapted from Fig. 3, Chapon \textit{et al.}\cite{chapon04}. Antiferromagnetically aligned moments lie on the $\mathrm{Mn^{3+}}$ (pink) and $\mathrm{Mn^{4+}}$ (green) ions that are located in square based pyramid and octahedral coordinations respectively; the terbium ions are shown in purple. Oxygen ions are omitted for clarity.}
\end{figure}

	At low temperature the manganese sublattice is found to exist in incommensurate (ICM) and commensurate (CM) magnetic phases, in which for both cases the manganese ion moments align in the \textit{ab} plane forming two spin density waves with wave vector ($\delta$, 0, $\tau$), relieving an otherwise geometrically frustrated system\cite{chapon04,blake05}. Along the \textit{c}-axis these moments alternate with ferromagnetic and antiferromagnetic layers\cite{blake05}. At $T_N=43\ \mathrm{K}$, the system enters an incommensurate phase (ICM2) in which manganese ions order antiferromagnetically with $\delta\simeq0.5$ and $\tau=0.3$\cite{chapon04}. Slightly lower in temperature, $T_{FE}=38\ \mathrm{K}$ marks the onset of ferroelectric order with $\mathbf{P}\parallel\mathit{b}$\cite{nhur}. The symmetry of the system in this ferroelectric phase is reduced, probably to space group $Pb\mathrm{2_1}m$\cite{koo:197601,blake05}. In the temperature range 33 K $>$ T $>$ 24 K  the system locks into a CM phase, ($\delta=\frac{1}{2}$, $\tau=\frac{1}{4}$), which is diagrammatically shown in Fig. \ref{magstruc}, as refined from neutron diffraction experiments.\cite{chapon04,blake05}  At T $<$ 24 K $\mathrm{TbMn_2O_5}$ enters another incommensurate phase (ICM1), ($\delta$ = 0.48, $\tau$ = 0.32)\cite{chapon04}, an unlikely transition as one might expect a commensurate ground state of any simple ordered system.
	
	Below $T_N$, the manganese magnetic structure spin polarizes the electronic states of terbium ions. A phenomena shown to occur and discussed not only in $\mathrm{TbMn_2O_5}$\cite{blake05}, but also in other manganese oxide multiferroics such as $\mathrm{TbMnO_3}$\cite{voigt:104431} and $\mathrm{HoMn_2O_5}$\cite{beutier-2008}. Blake \textit{et al.}\cite{blake05} suggested that this is specifically due to an interaction with the $\mathrm{Mn^{4+}}$ spin density. The existence of a separate magnetic order on the terbium ions was hypothesized with the terbium sublattice ordering below 10 K.\cite{saito95} In this paper we investigate the interaction with the manganese spins and any independent ordering of the terbium ion sublattice by resonant x-ray magnetic scattering (RXMS) full polarization analysis\cite{mazzoli:195118,detlefs:024425}. It can be assumed that the terbium 5d valence band has a large overlap not only with the lower energy terbium 4f band, but also with the 3d manganese band. The terbium 5d states will therefore be spin polarized via interaction with neighboring $\mathrm{Mn^{4+}}$, $\mathrm{Mn^{3+}}$, and Tb electronic spin density. At the terbium $\mathrm{L_{III}}$ edge, the 5d band is probed by the 2p $\Leftrightarrow$ 5d dipole transition. The 4f band of terbium, in which the unpaired electrons exist, is probed at this edge by the 2p $\Leftrightarrow$ 4f quadrupole transition.

\section{EXPERIMENT}\label{exp}
	
	A high quality single crystal of $\mathrm{TbMn_2O_5}$ with dimensions approximately 2 mm was grown at the Department of Chemistry of the National Taiwan University by flux growth. The single crystal was prepared such that the (4, 4, 0) Bragg reflection was close to the surface normal. The sample was mounted on the six-circle diffractometer in EH2 on beamline ID20 at the European Synchrotron Radiation Facility with the \textit{c}-axis perpendicular to the horizontal scattering plane. A helium flux cryostat was used to achieve a stable sample temperature through the different phases.
	
\begin{figure}
\includegraphics[width=7cm]{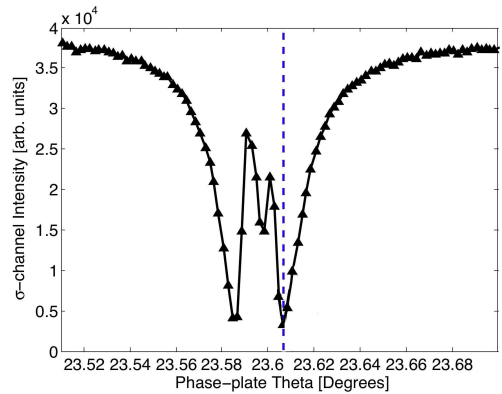}
\caption{\label{ppcal}(Color online) The transmitted intensity in the $\sigma$'-channel as a function of the phase plate theta angle. The diamond crystal behaves as a half-wave plate when aligned such that theta corresponds to a minimum in the graph as shown by the blue dashed line\cite{berman:1502}. This is symmetrical about the Bragg angle.}
\end{figure}

\begin{figure}
\includegraphics[width=8cm]{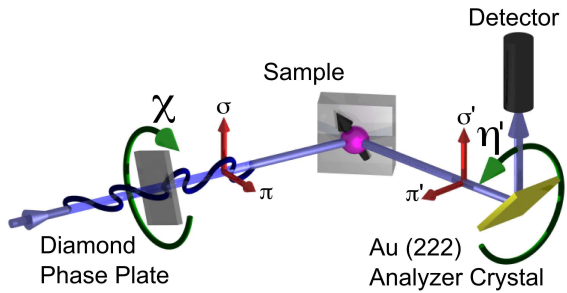}
\caption{\label{panalysis}(Color online) A schematic of the experimental setup. The polarized x-ray beam (blue) enters from the left. The polarization of the incident beam was rotated by a diamond phase-plate. The scattered beam polarization was determined through a secondary scattering process from an analyzer crystal prior to detection. Note the $\sigma$ and $\pi$ orientations where primes (eg. $\sigma$', $\pi$') refer to the scattered beam.}
\end{figure}

	The technique of using a phase plate, positioned in the incident beam such that it rotates the horizontally polarized light to any arbitrary linear polarization, was employed. A diamond crystal was used as the phase retarder by scattering near the (111) reflection. A convenient forward-scattering geometry is possible due to the low absorption coefficient of diamond. The phase plate was calibrated by measuring the scattered intensity in the $\sigma$ channel when rotating the phase plate theta through the Bragg angle, see Fig. \ref{ppcal}. A minimum in the calibration curve (blue dashed line in Fig. \ref{ppcal}) was selected as the angle at which the diamond crystal behaves as a half-wave plate. Note that a  $\chi$ phase plate rotation about the incident beam (see Fig. \ref{panalysis}) equates to a $\epsilon=2\chi$ rotation in polarization. The scattered beam polarization was determined using an Au(222) analyzer crystal. This technique of full polarization analysis is described in greater detail in the Appendix and is diagrammatically illustrated in Fig. \ref{panalysis}.

	The polarization of the scattered beam was measured as a function of incident polarization. By simulating the response and comparing it with the data, it was possible to refine the magnetic moment direction.

\section{RESULTS AND DISCUSSION}\label{results}

\begin{figure}
\includegraphics[width=8cm]{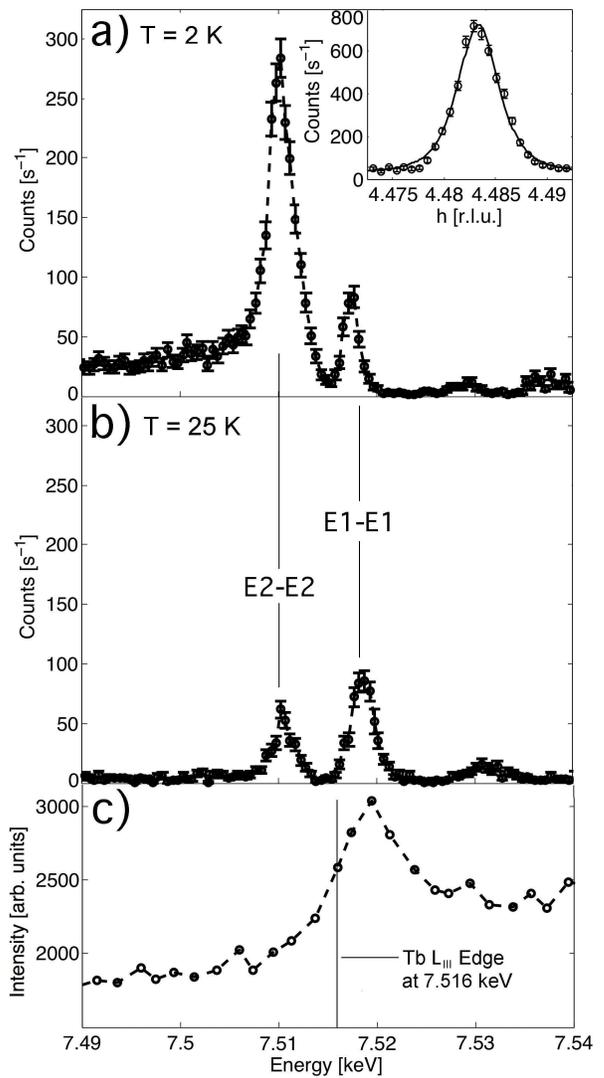}
\caption{\label{res}a) A scan of the scattered intensity of the (4.48, 4, -0.32) reflection as a function of the incident x-ray energy measured at constant wavevector at 2 K within the ICM1 phase, measured through the Tb $\mathrm{L_{III}}$ edge in the $\pi - \sigma'$ polarization channel. The inset shows the peak profile as measured along h. b) A similar scan of the (4.5, 4, -0.25) reflection measured at 25 K within the CM phase. c) The measured fluorescence curve of $\mathrm{TbMn_2O_5}$ showing the Tb $\mathrm{L_{III}}$ absorption edge at 7.516 keV.}
\end{figure}

	The first harmonic magnetic satellite reflection (4+$\delta$, 4, 0-$\tau$) was observed in the ICM1 phase (Fig. \ref{res}, inset). All the scattered intensity was found to be in the rotated $\pi-\sigma'$ channel, as expected of the magnetic scattering cross section. The peak, measured along the [100] direction, shows a good fit with a Lorentzian squared function centered at \textit{h} = 4.483. The inverse correlation length, defined as:
\begin{equation}
\xi^{-1}=\frac{2\pi}{\mathit{a}} \Delta \mathbf{h}_{HWHM},
\label{invcor}
\end{equation}
where \textit{a} is the lattice parameter and $\Delta\mathbf{h}$ is the half width at half maximum, gave $\xi^{-1}=2.212\times10^{-3} \pm6.2\times10^{-5} \mathrm{\AA^{-1}}$. The inverse correlation length of this reflection remained constant within experimental error within the temperature range 2$<$\textit{T}$<$35 K, except for a slight increase close to the ICM1/CM transition.

	Scans of the scattered intensity as a function of incident x-ray energy through the Tb $\mathrm{L_{III}}$ edge at constant wave vector were performed at the (4+$\delta$, 4, 0-$\tau$) reflection in both the incommensurate and commensurate phases. Two distinct energy resonances were observed, measured in the $\pi-\sigma'$ channel in both the low-temperature incommensurate and commensurate phases, and are shown in Fig. \ref{res}a and Fig. \ref{res}b, respectively. These excitations, which are evident in both phases, are centered at 7.510 and 7.518 keV. In comparison with the fluorescence spectrum as shown in Fig. \ref{res}c, the higher-energy peak occurs just above the Tb $\mathrm{L_{III}}$ absorption edge, as one would expect for an E1-E1 dipole transition. The other resonance is likely to be of quadrupolar E2-E2 origin. It exists 8 eV lower, a characteristic shift between dipole and quadrupole energy resonances in the rare-earth series\cite{PhysRevLett.79.3775, dallera2000, Wende02}. This resonance has been studied in magnetic field by Bland \textit{et al.}\cite{bland08}

\begin{figure}
\includegraphics[width=8cm]{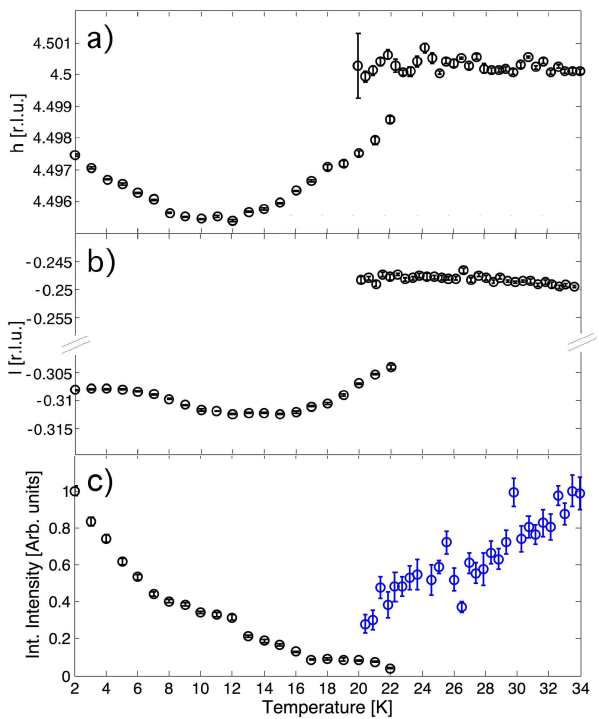}
\caption{\label{temp}(Color online) The wavevector (position in reciprocal space) measured along \textit{h} (a) and \textit{l} (b) of the (4+$\delta$, 4, 0-$\tau$) magnetic satellite reflection as a function of temperature in the range 2 K $<$ T $<$ 34 K. The CM - ICM1 phase transition at approximately 21 K is clearly seen, together with the phase coexistence occuring between 20 - 22 K. Within the ICM1 phase the incommensurate wavevector varies in both \textit{h} and \textit{l} as a function of temperature. c) The temperature dependence of the integrated intensity of the (4+$\delta$, 4, 0-$\tau$) reflection measured in the \textit{h} direction. The black (ICM1) and blue (CM) data were collected on different experiments. Direct comparison of the scattered intensity in the two phases is therefore not possible, however one can observe the trend of the CM reflection decreasing in intensity on decreasing temperature, and the growth of the ICM1 reflection on cooling below 22 K.}
\end{figure}

	The wave vector positions in reciprocal space and the transition temperatures of the ICM1 and CM reflections were measured and compared with neutron diffraction results to provide a good indication as to whether or not the reflections are of magnetic origin. Figure \ref{temp} shows the position in \textit{h} [Fig. \ref{temp}(a)] and position in \textit{l} [Fig. \ref{temp}(b)] as functions of temperature for the (4+$\delta$, 4, 0-$\tau$) peak. The transition from an incommensurate to a commensurate wave vector on warming is clear and agrees well with published $\delta$ and $\tau$ values\cite{chapon04,blake05}. In addition we observed that the two phases coexist in the temperature range of 20 - 22 K. This coexistence has also been observed by both soft x-ray and neutron diffraction studies\cite{kobayashi04,okamoto}, showing the phase transition to be strongly first order. The inverse correlation length as measured by the width of the reflection increases in the coexistence region, evidence of greater disorder of the terbium magnetic structure. Figure \ref{temp}c shows the temperature dependence of the integrated intensity of the (4+$\delta$, 4, 0-$\tau$) reflection measured in the \textit{h} direction. Upon cooling, the intensity of the CM reflection decreases. At 22 K, the ICM1 reflection simultaneously begins to increase in intensity and continues to do so below 20 K, at which temperature the CM reflection becomes extinct. The CM and ICM1 data, which are shown in the figure in blue and black, respectively, were measured on different experiments. It is therefore not possible to draw direct comparisons of the magnitude of the integrated intensity in the two phases.  We note that the widely acknowledged transition temperature of 24 K falls outside this range. This discrepancy is likely to be a result of x-ray beam heating or a discrepancy in the experimental calibration due to the limited proximity of the temperature sensor to the sample.

\begin{figure}
\includegraphics[width=8cm]{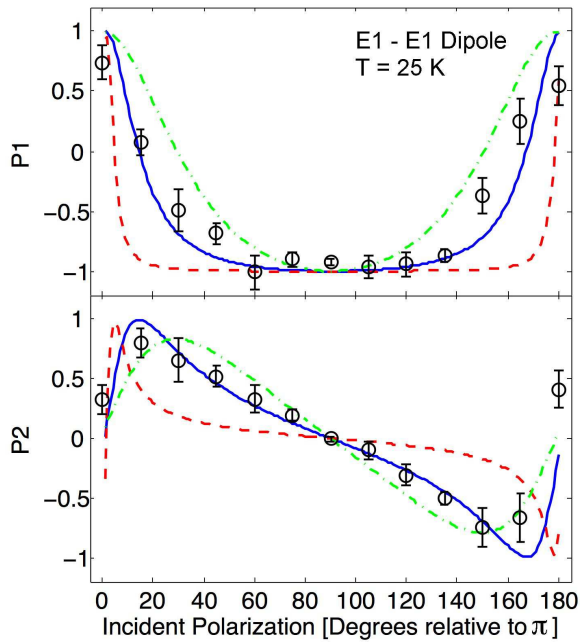}
\caption{\label{dip_com}(Color online) A plot of the measured Poincar$\mathrm{\acute{e}}$-Stokes parameters P1 and P2 as a function of the incident x-ray polarization of the commensurate (4+$\delta$, 4, 0-$\tau$) reflection at the E1-E1 energy resonance at 25 K. Simulations of the $\mathrm{Mn^{4+}}$ and $\mathrm{Mn^{3+}}$ magnetic structures [as refined by Blake \textit{et al.}\cite{blake05}] are shown as blue solid and red dashed lines respectively. A simulation of the terbium magnetic structure scattering at the E1-E1 transition in this phase, as refined later in this paper at the E2-E2 transition, is shown by the green dash-dotted line.}
\end{figure}

	Full polarization analysis of the (4+$\delta$, 4, 0-$\tau$) reflection was performed at an energy of 7.518 keV corresponding to the E1-E1 transition, probing the terbium 5d band. Figure \ref{dip_com} shows the Poincar$\mathrm{\acute{e}}$-Stokes parameters P1 and P2, which are defined as
\begin{subequations}\label{stokesparams}
\begin{align}
P1={}&(I_{\sigma'} -I_{\pi'})/(I_{\sigma'} +I_{\pi'})\\
P2={}&(I_{{+45^\circ}'}-I_{{-45^\circ}'})/(I_{{+45^\circ}'}+I_{{-45^\circ}'})
\end{align}
\end{subequations}	
measured as functions of incident polarization in the CM phase. It was hypothesized that the terbium ions are polarized by the manganese magnetic structure, in particular by the $\mathrm{Mn^{4+}}$ spin density\cite{blake05}.  As explained in the introduction, by probing the terbium 5d valence states at this transition, any refinement or simulation of magnetic structure would reveal the origin of the polarization. Superimposed in Fig. \ref{dip_com} are three simulations, the blue solid line shows the line shape expected if the terbium ion magnetism is solely a result of interaction with the $\mathrm{Mn^{4+}}$ spin density and the dashed red line is the line shape expected if the scattering occurs due to interactions with the $\mathrm{Mn^{3+}}$ spin density. The green dash-dotted line is the lineshape of a simulation of scattering due to interaction with terbium 4f magnetic moments, the directions of which were refined by the fit to the data taken at the E2-E2 transition presented later in this paper. This Tb 4f magnetic structure also exists as a consequence of spin polarization by the manganese magnetic structure and will itself interact with the 5d band. The reduced $\chi^2$ values of the $\mathrm{Mn^{4+}}$, $\mathrm{Mn^{3+}}$, and Tb simulations are 2.8, 17.9, and 5.8, respectively. This result conclusively verifies the hypothesis of Blake \textit{et al.}\cite{blake05} that the terbium sublattice is polarized by the close proximity of the $\mathrm{Mn^{4+}}$ spin density, as opposed to $\mathrm{Mn^{3+}}$ spin density. A reduced $\chi^2$ value of 2.8 of the $\mathrm{Mn^{4+}}$ simulation, when compared to the value of 5.8 of the Tb 4f band simulation, also shows that the Tb 5d band is predominantly polarized by the $\mathrm{Mn^{4+}}$ 3d band; however, one clearly cannot exclude the interaction with the terbium 4f spin configuration. Indeed, in a RXMS study of $\mathrm{HoMn_2O_5}$\cite{beutier-2008} performed at the dipole transition at the Ho L$\mathrm{_{III}}$ edge, an azimuthal dependence on scattered intensity measured in the commensurate phase, showed excellent agreement with a theoretical azimuthal dependence on scattering from holmium with the 5d band polarized by the 4f band magnetic structure.

\begin{figure}
\includegraphics[width=8cm]{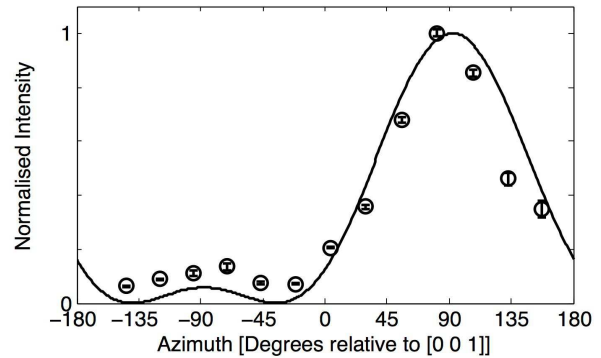}
\caption{\label{azi}The measured intensity of the commensurate (4+$\delta$, 4, 0+$\tau$) reflection at the E1-E1 energy resonance at 25 K in the $\sigma-\pi'$ channel as a function of azimuth angle of the sample. The solid line shows a simulation of the azimuthal dependence expected when the terbium magnetic ions are influenced solely by the $\mathrm{Mn^{4+}}$ magnetic structure [as refined by Blake \textit{et al.}\cite{blake05}].}
\end{figure}

	To confirm this result, a conventional azimuthal dependence of the scattered intensity in the $\sigma-\pi'$ channel was measured at the same energy at the (4+$\delta$, 4, 0+$\tau$) reflection in a vertical scattering geometry. The data, shown in Fig. \ref{azi}, compared to a simulation of the expected azimuthal dependence, assuming scattering from the terbium sublattice when polarized by $\mathrm{Mn^{4+}}$ spin density (solid line in Fig. \ref{azi}), also shows excellent agreement.

\begin{figure}
\includegraphics[width=8cm]{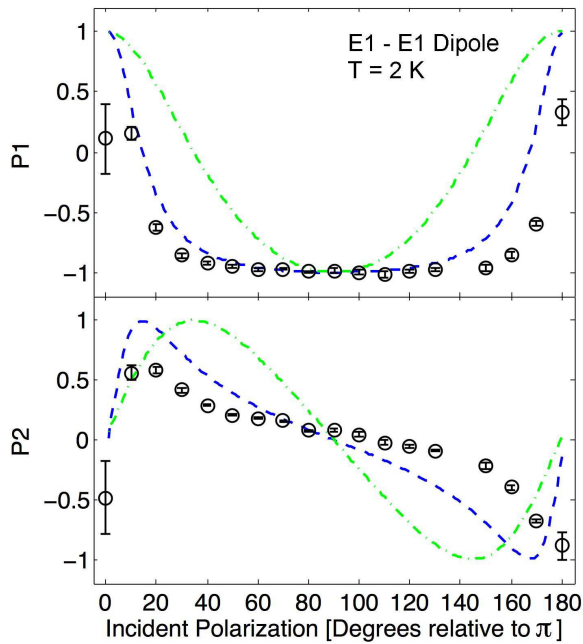}
\caption{\label{dip_incom}(Color online) A plot of the measured Poincar$\mathrm{\acute{e}}$-Stokes parameters P1 and P2 as a function of incident x-ray polarization of the incommensurate (4+$\delta$, 4, 0-$\tau$) reflection at the E1-E1 energy resonance at 2 K. The commensurate $\mathrm{Mn^{4+}}$ magnetic structure simulation shown in blue in Fig. \ref{dip_com} is shown here as a blue dashed line. The green dash-dotted line shows a simulation of the terbium magnetic structure scattering at the E1-E1 transition in the ICM1 phase, as refined later in this paper at the E2-E2 transition.}
\end{figure}

	The polarization analysis was performed again at the same energy (7.518 keV) in the incommensurate phase. The data is shown in Fig. \ref{dip_incom}. Due to a lack of a magnetic structure refinement in the literature of the ICM1 phase the large number of free parameters results in an unreliable refinement using this technique. We can therefore only make comparison to simulations of known magnetic structures. The CM $\mathrm{Mn^{4+}}$ magnetic structure simulation, which is shown to be of best agreement to the CM polarization analysis presented in Fig. \ref{dip_com}, is shown here as a blue dashed line. In comparison to the data, the sensitivity of this technique to a magnetic structure rearrangement upon the CM/ICM1 transition, resulting in a change of spin polarization of the 5d band, is clear. This could be due to a realignment of the manganese magnetic structure or a stronger interaction with the terbium sublattice or both. The green dash-dotted line plotted in Fig. \ref{dip_incom} is a simulation of the scattering expected due to interaction of the Tb 5d band with the terbium magnetic structure, as refined later in this paper at the E2-E2 transition. From the poor agreement of the terbium simulation, we would conclude that in this phase the 5d band remains predominantly polarized by the manganese magnetic structure. The difference in full polarization analysis data between the two phases is therefore likely be a result of a realignment of $\mathrm{Mn^{4+}}$ moments.

\begin{figure}
\includegraphics[width=8cm]{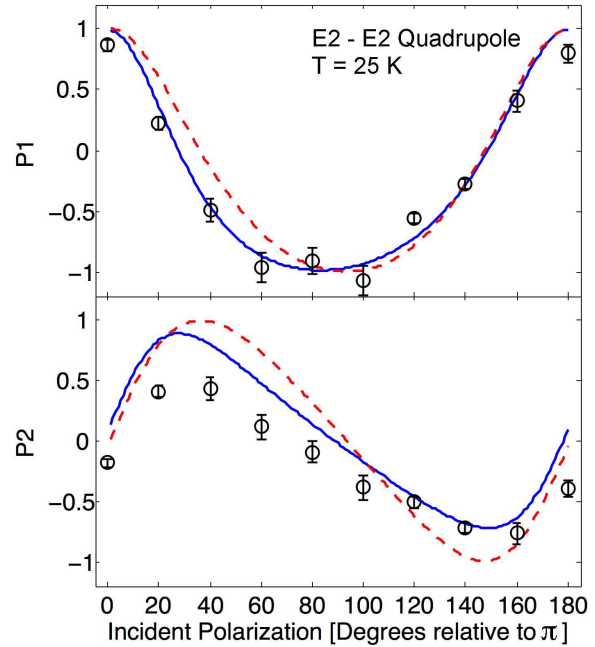}
\caption{\label{quad_com}(Color online) A plot of the measured Poincar$\mathrm{\acute{e}}$-Stokes parameters P1 and P2 as a function of incident x-ray polarization of the commensurate (4+$\delta$, 4, 0-$\tau$) reflection at the E2-E2 energy resonance at 25 K. A least squares fit with a reduced $\chi^2$ value of 13.9 is shown in blue (solid line). See text for the refined moment directions. The incommensurate fit performed in Fig. \ref{quad_incom} is shown by the red dashed line.}
\end{figure}

\begin{figure}
\includegraphics[width=7cm]{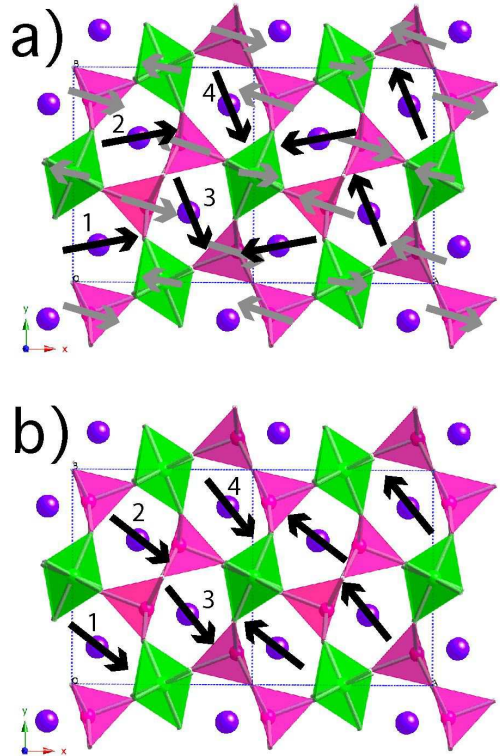}
\caption{\label{tbmag}(Color online) Diagrammatic illustration of the refined terbium ion magnetic moment directions (black arrows) in a) the CM phase and b) the ICM1 phase. There exists a slight component out of the \textit{ab} plane not shown here. The manganese ion magnetic moment directions in the commensurate phase are shown by grey arrows as in figure \ref{magstruc}. These directions are currently unknown in the incommensurate phase.}
\end{figure}

	The polarization analysis measurement was conducted in the commensurate phase at the E2-E2 resonance (7.510 keV). At this transition we directly probe unpaired electrons in the terbium 4f band and, hence, the terbium sublattice magnetic structure. Figure \ref{quad_com} shows the data and a least-squares fit (blue solid line), similar to the analysis used in a study of $\mathrm{UPd_3}$\cite{walker:137203}. There are four crystallographically distinct terbium ions in the magnetic unit cell, labeled here according to the convention used by Blake \textit{et al.},\cite{blake05} as shown in Fig. \ref{tbmag}. In a recent neutron single crystal diffraction study of $\mathrm{HoMn_2O_5}$\cite{vecchini:134434}, which shows very similar macroscopic properties and microscopic structure to $\mathrm{TbMn_2O_5}$, holmium ion positions 1 and 2 were found to have the same moment direction as were holmium ions 3 and 4. This is in contradiction to the earlier neutron powder diffraction study by Blake \textit{et al.}\cite{blake05} in which both Tb and Ho compounds have the rare-earth positions 1-4 and 2-3 paired in the refinement. In fitting the data it was evident that it was necessary to pair the terbium ions as in the recent $\mathrm{HoMn_2O_5}$ single crystal study. Fixing the \textit{a} component of moments on sites 1 and 2 and the \textit{b} component of moments on sites 3 and 4 gave a fit with four degrees of freedom. Magnetic moment directions on the terbium 1 and 2 sites were refined to be $10.5\pm2.6^\circ$ in the \textit{ab}-plane relative to the \textit{a}-axis and $0.2^\pm0.1\circ$ out of plane, and on the terbium 3 and 4 sites $292.5^\pm2.0\circ$ in the \textit{ab}-plane relative to the \textit{a}-axis and $0.2^\pm0.1\circ$ out of plane.
	
	These moment directions are in contradiction with neutron powder diffraction measurements on the $R\mathrm{Mn_2O_5}$ series as reported by Blake \textit{et al.},\cite{blake05} however, so are recent refinements of $\mathrm{HoMn_2O_5}$ by neutron and synchrotron single-crystal data\cite{vecchini:134434,beutier-2008,kimura07} when compared to the same 2005 study. In these $\mathrm{HoMn_2O_5}$ experiments, holmium directions were reported to be pointing along the \textit{a} and \textit{b} axes with a small component in the direction of the \textit{c}-axis. These directions are comparable to those refined in this study on $\mathrm{TbMn_2O_5}$. Here, terbium moments are found to be coupled in the same fashion and lie approximately along the crystallographic axes in the \textit{ab}-plane (the moments are in fact refined to be canted away from the axes by approximately $10^\circ$) with a slight \textit{c}-axis component. However, the configuration of the spin orientations of the rare earth and the $\mathrm{Mn^{4+}}$ magnetic structure differ to that of the holmium compound. As a consequence, it is not possible to explain the terbium ordering in terms of a simple antiferromagnetic superexchange interaction between terbium and nearest-neighbor $\mathrm{Mn^{4+}}$ ions where the manganese moment \textit{a} and \textit{b} axis components cancel at sites 1-2 and 3-4, respectively, the argument which was put forward by C. Vecchini \textit{et al.}\cite{vecchini:134434} in the $\mathrm{HoMn_2O_5}$ study.
	
	The ICM1 phase of the $R\mathrm{Mn_2O_5}$ series is less understood, with no complete neutron or synchrotron study of the magnetic structure published to date. Full polarization analysis performed below 10 K at the E2-E2 transition will refine the moment directions of the terbium ion magnetic order.\cite{nhur} These can then be compared to those refined in the commensurate phase.

\begin{figure}
\includegraphics[width=8cm]{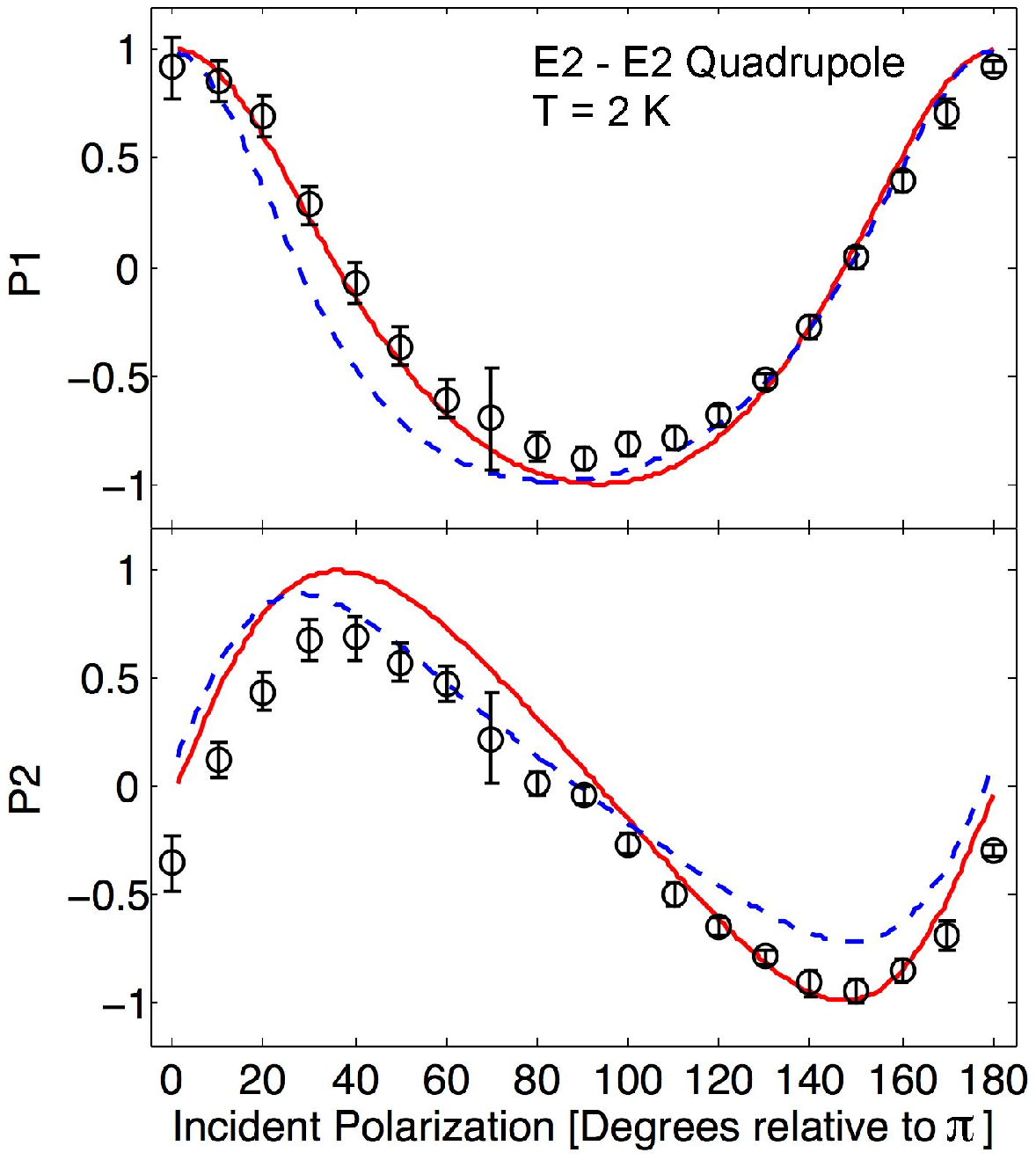}
\caption{\label{quad_incom}(Color online) A plot of the measured Poincar$\mathrm{\acute{e}}$-Stokes parameters P1 and P2 as a function of incident x-ray polarization of the incommensurate (4+$\delta$, 4, 0-$\tau$) reflection at the E2-E2 energy resonance at 2 K.  A least squares fit with a reduced $\chi^2$ value of 7.5 is shown in red (solid line). See text for the refined moment directions. The commensurate fit performed in Fig. \ref{quad_com} is shown by the blue dashed line.}
\end{figure}

	Figure \ref{quad_incom} shows the Poincar$\mathrm{\acute{e}}$-Stokes parameters P1 and P2 measured as a function of incident x-ray polarization at 2 K in the ICM1 phase. Superimposed is a least-squares fit (red solid line), assuming the same pairing of terbium as in the CM phase. Again the \textit{a} and \textit{b} components of moments on sites 1-2 and 3-4, respectively, were fixed, giving a fit with four degrees of freedom. The refined moment directions on the terbium 1 and 2 sites were $322.0^\pm4.6\circ$ in the \textit{ab}-plane relative to the \textit{a}-axis and $0.5^\pm0.1\circ$ out of plane, and on the terbium 3 and 4 sites it was $308.9^\pm10.4\circ$ in the \textit{ab}-plane relative to the \textit{a}-axis and $0.6^\pm0.1\circ$ out of plane. This refinement of the 2 K terbium sublattice magnetic structure results from Tb-Tb ordering below 10 K as predicted\cite{nhur}, as well as interaction with the $\mathrm{Mn^{4+}}$ sublattice, the magnetic structure of which is unknown in the ICM1 phase. However below 10 K, the terbium order dominates the bulk magnetization of the sample\cite{nhur}.
	
	Also plotted as dashed lines in Fig. \ref{quad_com} (CM) and Fig. \ref{quad_incom} (ICM1) are the ICM1 and CM fits, respectively. On visual inspection there appears to be little difference between the two fits, particularly when comparing them with the data in Fig. \ref{quad_incom}. It is surprising then to find that the refinement suggests significantly different spin configurations. The fitting process is clearly sensitive to small changes in the the polarization dependence and therefore demands high experimental accuracy. This poses an important limitation of the technique, one that must be considered when interpreting the results. However, due to the small error on the data, we have refined the moment directions with a high degree of precision.

	The $\mathrm{Mn^{4+}}$ and $\mathrm{Mn^{3+}}$ magnetic structures have been shown not to change on substitution of different $R^{3+}$ rare-earth ions in the commensurate phase.\cite{kimura07} This strongly suggests that the manganese ion magnetic structures are responsible for the magnetoelectric coupling in the $\mathit{R}\mathrm{Mn_2O_5}$ series. Experimental confirmation that the terbium ions are spin polarized specifically by the $\mathrm{Mn^{4+}}$ spin density is therefore of great significance in further understanding of the underlying coupling mechanism, and, by extension, the magneto-electric properties of these compounds.
	
	Probing the terbium ion 4f band in the CM and ICM1 phases at 25 and 2 K, respectively, has provided a direct measurement of the terbium sublattice spin configuration in both phases. Importantly, at 2 K we have observed the terbium sublattice ordering itself. By assuming a pairing of terbium ions as found by neutron single-crystal data on $\mathrm{HoMn_2O_5}$,\cite{vecchini:134434} precise moment directions of the probable terbium magnetic structures were found. We observed a change in spin configuration due to a weakening of the terbium ion average magnetization as expected above 10 K,\cite{nhur} allowing for a stronger interaction with the $\mathrm{Mn^{4+}}$ spin density, as shown by the dipolar resonance that probes the 5d valence band. Figure \ref{tbmag} diagrammatically illustrates the terbium moment directions refined in both the CM [Fig. \ref{tbmag}(a)] and the ICM1 [Fig. \ref{tbmag}(b)] phases. The moment directions on manganese ions have been omitted in Fig. \ref{tbmag}b as they are unknown in the incommensurate phase. By visual comparison of the polarization dependence we observe little change in CM/ICM1 transition; however, when fitting to the data we refine a significant change in spin configuration with precise error bounds. As previously discussed, this illustrates a key limitation of the technique.

\section{CONCLUSIONS}
	Two energy resonances have been experimentally observed at the (4+$\delta$, 4, 0-$\tau$) reflection, one at the Tb $\mathrm{L_{III}}$ edge while the other at 8 eV lower, in both the low-temperature (ICM1) and (CM) phases of $\mathrm{TbMn_2O_5}$. The higher energy resonance originates from the E1-E1 dipolar transition, which probes the terbium 5d band, and the lower energy resonance originates from the E2-E2 quadrupolar transition, which probes the terbium 4f band. The temperature dependence and wave vector measurements confirmed the reflection to be of magnetic origin. Full polarization analysis conducted at the E1-E1 resonance confirmed that the 5d band of terbium is polarized predominantly by the $\mathrm{Mn^{4+}}$ spin density in the commensurate phase, with an additional interaction with the spin polarized Tb 4f magnetic structure. Tuning to the E2-E2 excitation at 2 K (ICM1) and fitting to full polarization analysis data enabled the refinement of directions of the spin ordering of the unpaired terbium 4f electrons. The measurement and refinement were repeated in the CM phase.

\begin{acknowledgments}
	We would like to acknowledge Prof. Ru-Shi Liu for the crystal growth. R.D.J., S.R.B. and T.A.W.B. would like to thank S.T.F.C. and EPSRC for the funding. We are grateful to the European Synchrotron Radiation Facility for the beamtime and access to their facilities. The work at Brookhaven National Laboratory is supported by the Office of Science, U.S. Department of Energy, under Contract No. DE-AC02-98CH10886.
\end{acknowledgments}

\appendix*
\section{FULL POLARIZATION ANALYSIS}\label{theory}

	The scattering amplitude, $f^{RXMS}$, of resonant x-ray magnetic scattering is anisotropic; dependent upon the direction of magnetic moment. It contains components from both the dipole, E1-E1 (see Eq. \ref{e1sfactor}), and the typically less intense quadrupole, E2-E2 (see Eq. \ref{e2sfactor}), transitions. The E2-E2 transition usually occurs a few eV below the absorption edge, due to a stronger interaction between the core hole and excited electron.\cite{detlefs97} The structure factors are extensively derived from spherical harmonics by Blume and Gibbs,\cite{PhysRevB.37.1779} Hannon \textit{et al.}\cite{hannonrxs}, and Hill and McMorrow\cite{HILL:1996lr}. The theory presented here is minimal compared to that described in the aforementioned original literature. Only the final results useful in data analysis are given, that is, the scattering amplitude expressed in terms of polarization and magnetic moment direction.  The formulation of the following equations\cite{hannonrxs,HILL:1996lr} (Eqs. \ref{e1sfactor} and \ref{e2sfactor}) is based on the assumption that the system is isotropic where only the electron spin breaks the symmetry. In the case of magnetically frustrated multiferroics, of which $\mathrm{TbMn_2O_5}$ is an example, this assumption is an over simplification. However the scattering amplitude in the form presented here has been successfully employed, in particular, to the closely related compound $\mathrm{TbMnO_3}$\cite{voigt:104431}. There are three terms that contribute to the E1-E1 resonant scattering amplitude and five terms that contribute to the E2-E2. In this experiment, we measured first harmonic magnetic satellite reflections and are therefore only concerned with the first order terms.\cite{hannonrxs,HILL:1996lr} At the E1-E1 transition,
\begin{equation}
f^{RXMS}_{E1-E1}=-iF_{E1-E1}^{(1)}(\hat{\epsilon}'\times\hat{\epsilon})\cdot\hat{\mathbf{z}}
\label{e1sfactor}
\end{equation}
and at the E2-E2 transition,
\begin{equation}\label{e2sfactor}
\begin{split}
f^{RXMS}_{E2-E2}={}&-iF^{(1)}_{E2-E2}[(\hat{\mathbf{k}}'\cdot\hat{\mathbf{k}})(\hat{\epsilon}'\times\hat{\epsilon})\cdot\hat{\mathbf{z}}  \\
&+(\hat{\mathbf{k}}'\times\hat{\mathbf{k}})(\hat{\epsilon}'\cdot\hat{\epsilon})\cdot\hat{\mathbf{z}}] \\
\end{split}
\end{equation}
where $\hat{\mathbf{k}}$ is the x-ray wave vector, $\hat{\epsilon}$ is the x-ray linear polarization orientation, and $\hat{\mathbf{z}}$ is the unit vector in the direction of the ion's magnetic moment. $F_{E1-E1}^{(1)}$ and $F^{(1)}_{E2-E2}$ are coefficients dependent upon the transition in question, as defined by Hannon \textit{et al.}\cite{hannonrxs} and Hill and McMorrow\cite{HILL:1996lr}, that determine the strength of the resonance. Primes refer to the scattered beam. The polarization, magnetic moment and scattering amplitudes are defined using the common coordinate system as defined by Blume and Gibbs\cite{PhysRevB.37.1779}(see Eqs. \ref{p1p2where}).

	The total scattering amplitude incorporates a phase factor and is written as
\begin{equation}
\mathbf{F}=\sum_jf^{RXMS}_{E1,E2}e^{i\mathbf{q}.\mathbf{r_j}}
\label{tsfactor}
\end{equation}
where \textbf{q} is the scattering vector and \textbf{r} is the crystallographic coordinate of the $\mathrm{jth}$ terbium ion.
	
	The incident x-ray polarization was rotated by a diamond phase plate. Dynamical theory shows that birefringence occurs in perfect crystals when scattering at, or near, a Bragg reflection. Experimentally, the polarization ellipticity and handedness induced by the birefringence is determined by the thickness of the crystal and the deviance from the Bragg angle.\cite{Giles:1995lr,bouchenoire03} The efficiency of the phase plate is therefore limited by the beam divergence. Far from the Bragg condition the incident polarization is unrotated. As the Bragg condition is approached, a phase difference between pi and sigma channels relative to the crystal is induced and elliptically polarized light is produced. At a particular deviation a $\pi$ phase difference is induced and the phase plate behaves as a half wave plate. By maintaining this deviation from the Bragg condition and rotating on a $\chi$-circle about the beam, any orientation of linearly polarized light can be selected.\cite{mazzoli:195118}
		
	The polarization of the scattered beam was analyzed by a crystal selected and cut such that the beam Bragg diffracts with minimum absorption close to Brewster's angle of 45$^\circ$; hence, only scattering x-rays with incident polarization perpendicular to the analyzer scattering plane. Therefore by rotating the analyzer crystal about the scattered beam by an angle $\eta'$, a polarization selective measurement can be made. Poincar$\mathrm{\acute{e}}$-Stokes parameters were determined by fitting the integrated intensity of the analyzer rocking curve as a function of $\eta'$ to the following equation;\cite{mazzoli:195118}
\begin{equation}
I=\frac{I_0}{2}[1+P1\cos2\eta'+P2\sin2\eta'],
\label{stokescalc}
\end{equation}
where $\eta' = 0^\circ$ and  $\eta' = 90^\circ$ correspond to polarization perpendicular, $\sigma'$, and parallel, $\pi'$ to the scattering plane, respectively.

	By writing P1 and P2 using Eqs. \ref{stokesparams} and \ref{tsfactor},
	\begin{widetext}		
\begin{equation}	
	P1=\frac{\lvert\sum_jf^{RXMS}_{E1,E2}(\hat{\epsilon}'_\sigma,\hat{\epsilon},\hat{\mathbf{z}}_j,\hat{\mathbf{k}}',\hat{\mathbf{k}})e^{i\mathbf{q}.\mathbf{r_j}}\rvert^2 - \lvert \sum_jf^{RXMS}_{E1,E2}(\hat{\epsilon}'_\pi,\hat{\epsilon},\hat{\mathbf{z}}_j,\hat{\mathbf{k}}',\hat{\mathbf{k}})e^{i\mathbf{q}.\mathbf{r_j}}\rvert^2}{\lvert \sum_jf^{RXMS}_{E1,E2}(\hat{\epsilon}'_\sigma,\hat{\epsilon},\hat{\mathbf{z}}_j,\hat{\mathbf{k}}',\hat{\mathbf{k}})e^{i\mathbf{q}.\mathbf{r_j}}\rvert^2 + \lvert \sum_jf^{RXMS}_{E1,E2}(\hat{\epsilon}'_\pi,\hat{\epsilon},\hat{\mathbf{z}}_j,\hat{\mathbf{k}}',\hat{\mathbf{k}})e^{i\mathbf{q}.\mathbf{r_j}}\rvert^2}
\label{p1actual}	
\end{equation}
and
\begin{equation}	
	P2=\frac{\lvert \sum_jf^{RXMS}_{E1,E2}(\hat{\epsilon}'_{+45^\circ},\hat{\epsilon},\hat{\mathbf{z}}_j,\hat{\mathbf{k}}',\hat{\mathbf{k}})e^{i\mathbf{q}.\mathbf{r_j}}\rvert^2 - \lvert \sum_jf^{RXMS}_{E1,E2}(\hat{\epsilon}'_{-45^\circ},\hat{\epsilon},\hat{\mathbf{z}}_j,\hat{\mathbf{k}}',\hat{\mathbf{k}})e^{i\mathbf{q}.\mathbf{r_j}}\rvert^2}{\lvert \sum_jf^{RXMS}_{E1,E2}(\hat{\epsilon}'_{+45^\circ},\hat{\epsilon},\hat{\mathbf{z}}_j,\hat{\mathbf{k}}',\hat{\mathbf{k}})e^{i\mathbf{q}.\mathbf{r_j}}\rvert^2 + \lvert \sum_jf^{RXMS}_{E1,E2}(\hat{\epsilon}'_{-45^\circ},\hat{\epsilon},\hat{\mathbf{z}}_j,\hat{\mathbf{k}}',\hat{\mathbf{k}})e^{i\mathbf{q}.\mathbf{r_j}}\rvert^2}
\label{p2actual}	
\end{equation}
\end{widetext}
where, as defined by Blume and Gibbs,\cite{PhysRevB.37.1779}
\begin{subequations}\label{p1p2where}
\begin{align}
\hat{\mathbf{z}}_j{}&=a_j\ \mathbf{\hat{u}_1}+b_j\ \mathbf{\hat{u}_2}+c_j\ \mathbf{\hat{u}_3}\\
\hat{\epsilon}'_\sigma{}&=\mathbf{\hat{u}_2}\\
\hat{\epsilon}'_\pi{}&=-\sin\theta\ \mathbf{\hat{u}_1}-\cos\theta\ \mathbf{\hat{u}_3}\\
\hat{\epsilon}'_{+45^\circ}{}&=\frac{1}{\sqrt{2}}[-\sin\theta\ \mathbf{\hat{u}_1}+\mathbf{\hat{u}_2}-\cos\theta\ \mathbf{\hat{u}_3}]\\
\hat{\epsilon}'_{-45^\circ}{}&=\frac{1}{\sqrt{2}}[\sin\theta\ \mathbf{\hat{u}_1}+\mathbf{\hat{u}_2}+\cos\theta\ \mathbf{\hat{u}_3}]\\
\hat{\epsilon}{}&=\sin(2\chi)\sin\theta\ \mathbf{\hat{u}_1}+\cos(2\chi)\ \mathbf{\hat{u}_2} -\sin(2\chi)\cos\theta\ \mathbf{\hat{u}_3}\\
\hat{\mathbf{k}}'{}&=\cos\theta\ \mathbf{\hat{u}_1}-\sin\theta\ \mathbf{\hat{u}_3}\\
\hat{\mathbf{k}}{}&=\cos\theta\ \mathbf{\hat{u}_1}+\sin\theta\ \mathbf{\hat{u}_3}
\end{align}
\end{subequations}	
simulations of different magnetic moment directions and scattering factors can be made. By using a least-squares method, where $a_j$, $b_j$, and $c_j$ are free parameters, it is possible to refine the magnetic moment direction from a measurement of P1 and P2 as a function of incident polarization.

\bibliography{multiferroic2}

\end{document}